\documentclass[aip,graphicx]{revtex4-1}
\usepackage{graphics,color}
\usepackage{graphicx}         
\usepackage{textcomp}

\draft % marks overfull lines with a black rule on the right

\begin{document}

%\preprint{}

\title{Seeing the vibrational breathing of a single molecule through time-resolved coherent anti-Stokes Raman scattering } %Title of paper



\author{Steven Yampolsky}
\affiliation{Department of Chemistry, University of California, Irvine, CA 92697, USA}

\author{Dmitry A. Fishman}
\affiliation{Department of Chemistry, University of California, Irvine, CA 92697, USA}

\author{Shirshendu Dey}
\affiliation{Department of Chemistry, University of California, Irvine, CA 92697, USA}

\author{Eero Hulkko}
\altaffiliation{Nanoscience Center, Department of Chemistry, P. O. Box 35, FI-40014, University of Jyv\"askyl\"a, Finland}

\author{Mayukh Banik}
\affiliation{Department of Chemistry, University of California, Irvine, CA 92697, USA}

\author{Eric O. Potma}
\affiliation{Department of Chemistry, University of California, Irvine, CA 92697, USA}
\email[]{epotma@uci.edu}

\author{Vartkess A. Apkarian}
\affiliation{Department of Chemistry, University of California, Irvine, CA 92697, USA}
\email[]{aapkaria@uci.edu}

\date{\today}

\begin{abstract}
The motion of chemical bonds within molecules can be observed in real time, in the form of vibrational wavepackets prepared and interrogated through ultrafast nonlinear spectroscopy. Such nonlinear optical measurements are commonly performed on large ensembles of molecules, and as such, are limited to the extent that ensemble coherence can be maintained. Here, we describe vibrational wavepacket motion on single molecules, recorded through time-resolved, surface-enhanced, coherent anti-Stokes Raman scattering. The required sensitivity to detect the motion of a single molecule, under ambient conditions, is achieved by equipping the molecule with a dipolar nano-antenna (a gold dumbbell). In contrast with measurements in ensembles, the vibrational coherence on a single molecule does not dephase. It develops phase fluctuations with characteristic statistics. We present the time evolution of discretely sampled statistical states, and highlight the unique information content in the characteristic, early-time probability distribution function of the signal.
\end{abstract}


\maketitle %\maketitle must follow title, authors, abstract and \pacs

%\section{test}
%\label{}
%\subsection{test}
%\subsubsection{test}
The demonstration of optically detected single molecules under cryogenic conditions, through absorption\cite{Moerner1989} and soon afterward through fluorescence\cite{Orrit1990}, can be recognized as the nascence of single molecule spectroscopy (SMS). The field was further propelled by demonstrations of SMS under ambient conditions, through near-field\cite{Betzig1993} and far-field optical microscopy.\cite{Nie1994, Eigen1994} The most common approach in SMS is the use of chromophores that strongly couple to light, to detect light-matter interactions through a variety of methods:\cite{Kulzer2004} absorption,\cite{Moerner1989, Kukura2010} resonant Raman,\cite{Moerner2002} fluorescence induced through linear\cite{Orrit1990} or nonlinear absorption,\cite{Plakhotnik1996} photo-thermal effect,\cite{Gaiduk2010} being examples. The principle aim of SMS is to interrogate individual molecular properties rather than averages of the vast ensembles typically probed in molecular spectroscopy. Ensemble averaging eliminates outliers, such as the tail of the Boltzmann distribution where most of chemistry hides.

The incoherent spectroscopy of chromophores mainly addresses environmental fluctuations that occur on timescales longer than the measurement time.\cite{Barkai2004} On timescales of ms and longer over which signals are accumulated, SMS is well matched to track bio-molecular transformations, such as enzymatic activity and protein folding.\cite{Weiss1999, Xie2001} The femtosecond-picosecond scale over which bonds move, break and form, poses a greater challenge. An important advance in this regard is the recent series of ultrafast pump-probe measurements using phase-locked pulse pairs,\cite{Hildner2011a, Brinks2011, Brinks2010, Hildner2011b} in which vibrational wavepackets and coherent manipulation of quantum bits were demonstrated on single molecules.  The information content of these measurements, however, is limited by electronic dephasing that occurs on timescales shorter than vibrational periods of motion. This characteristic of chromophores is implicit in their structureless and broad absorption spectra.

Raman spectroscopy avoids electronic dephasing considerations since it does not require evolution in real electronic states. As such, time-resolved coherent Raman spectroscopy is ideally suited to capture the motion of molecules in their ground electronic state. Although Raman is a feeble effect, it can be greatly amplified through plasmonic antennae,\cite{Novotny2011} namely through the surface enhanced Raman scattering (SERS) effect.\cite{Jeanmaire1977, Schatz1984} Following the initial demonstrations of single molecule sensitivity,\cite{Nie1997, Kneipp1997} SM-SERS has rapidly developed\cite{Etchegoin2008}, with ultimate sub-molecular spatial resolution recently achieved at the plasmonic junction of a scanning tunneling microscope.\cite{Zhang2013} 

The sensitivity in SM-SERS derives from enhanced local fields concentrated at junctions of metallic nano-structures. It is not clear whether the principles of SERS can be directly translated to the ultrafast time domain. Molecules subject to field gradients may migrate due to optical forces, and at incident intensities of 10$^{10}$ W/cm$^2$ reached in ultrashort laser pulses, nanoparticles melt, change shape and even fragment.\cite{Link2000} Surface-enhanced coherent anti-Stokes Raman scattering\cite{Voronine2012, Steuwe2011, Namboodiri2011, Koo2005, Ichimura2004, Chen1979} and stimulated Raman scattering\cite{Frontiera2011} has been demonstrated using fs lasers in  frequency domain measurements on SERS-active ensembles. And although a time-resolved, surface enhanced Raman Scattering (tr-SECARS) measurement has been reported in a colloidal ensemble, the unique capabilities of coherent Raman techniques to capture the evolution of vibrational wave packets\cite{Mukamel1995, Karavitis2001} has yet to be demonstrated in the single molecule limit.  Here, we report this realization. 

Through tr-CARS studies on individual nano-dumbbells, we capture the vibrational motion of single molecules in real-time, and we highlight the unique information content of time-resolved measurements in the single molecule limit. Rather than pure dephasing, coherences develop phase noise due to discrete sampling statistics, familiar from quantum optics.\cite{Loudon2000} We show that the probability distribution function (PDF) of the noise accumulated during early time evolution can be used to uniquely distinguish between one, versus two, versus many-molecule response. 

The experiments are carried out on trans 1,2-\textit{bis}-(4-pyridyl) ethylene (BPE) molecules attached to gold nano-dumbbells encapsulated in porous silica shells (Fig. 1a). That single molecule sensitivity can be reached through SERS at the junction between two gold nano-spheres, has been catalogued in the literature on this and  related systems\cite{Kleinman2012, Wustholz2010, Whitmore2011} including the direct isotopologue test of single molecule response.\cite{Dieringer2007, Wang2013} We do not have direct knowledge of the location or coverage of the molecules, which are adsorbed on the gold spheres prior to encapsulation. However, signatures of single molecule behavior are evident in the spectral fluctuations observed in sequentially recorded SERS spectra on individual nanostructures. We see meandering Raman trajectories, dramatic intensity fluctuations, and the appearance of lines that cannot be explained through dipolar Raman alone (Supplementary Information). These observations parallel a prior study on silver nano-dumbbells, where it was shown that the tensor nature of Raman scattering is in full force in SM-SERS and that spectral variations accompany changes in the relative orientation between molecule and local field, and the latter evolve due to structural changes of the irradiated junction.\cite{Banik2012} 

The spectrum shown in Fig. 1 is from a relatively stationary SERS trajectory, obtained at an incident intensity of 30 $\mu$W/$\mu$m$^2$. Remarkably, the SERS linewidths are significantly broader than those in the ensemble spectrum of solid BPE. We will focus on the strong C=C stretching modes near 1600 cm$^{-1}$ which arise from mixing between the ethylinic stretch and the pyridine ring breathing modes of BPE.\cite{Yang1996}  These lines have Lorentzian profiles with FWHM of 5 cm$^{-1}$ in solid BPE, but appear as Gaussians of FWHM $\sim$15 cm$^{-1}$ on the dumbbell (Fig.1d). SERS trajectories recorded at a rate of 1 spectrum/2s show resolution-limited lines of FWHM = 4 cm$^{-1}$ inside the envelope (Supplementary Information). Consistent with recent tip enhanced Raman measurements at cryogenic temperatures, it is clear that intrinsic SERS linewidths (limited by vibrational dissipation) are much sharper. Evidently, the broad Gaussian envelopes represent inhomogeneous distributions sampled by spectral diffusion, and both optical forces and local heating can be expected to contribute to the observed spectral distribution. The latter effect appears to be greatly reduced under pulsed irradiation conditions of the ultrafast tr-CARS measurements.

The tr-CARS measurements are carried out on drop cast particles on a 15 nm thick silicon nitride substrate. After SEM mapping and SERS characterization, the particles are transferred to a femtosecond laser scanning CARS microscope. The imaging system is interfaced with a tunable femtosecond laser (80 MHz repetition rate), which provides the pump ($\lambda_{pu} = 714$ nm), Stokes ($\lambda_{st} = 809$ nm) and probe ($\lambda_{pr} = 714$ nm) pulses for the electronically non-resonant CARS process illustrated in Figure 2a. The difference frequency between the coincident pump and Stokes pulses, $\omega_{pu}-\omega_{st} = 1640$ cm$^{-1}$, is selected to excite the superposition of bright modes that fall under the spectral bandwidth (Fig. 2a), to prepare the second order wavepacket, $\phi^{(2)}=\sum_\nu a_\nu\left|\nu\right>$. The evolving vibrational coherence $\left|\phi^{(2)}(t)\right>\left<\phi^{(0)}(t)\right|$ is then interrogated with the time-delayed probe pulse, and detected through the anti-Stokes shifted photons. Care is taken to minimize the average illumination intensity to below 20 $\mu$W/$\mu$m$^2$ to avoid photo-damage. 
 
The CARS image of a distribution of particles is presented in Fig. 2b. The nanoparticles are readily imaged through the electronic CARS response of the metal surface plasmon.\cite{Danckwerts2007, Wang2011} The variation in their intensities is readily associated with the distribution of orientations relative to the linear polarization of the excitation. The molecular CARS response is distinguished by its dependence on the delay of the probe pulse and vibrational resonance (Fig. 2c,d). Beyond 0.2 ps, the instantaneous electronic response of the plasmon reduces to a constant background over which the modulated molecular contribution appears. The spectrally resolved tr-CARS signals illustrate that when the difference frequency, $\omega_{pu}-\omega_{st} = 1800$ cm$^{-1}$, is detuned from the molecular resonance, only the constant background of two-beam electronic CARS remains. On resonance, a periodically modulated signal is seen. The period, $\tau\sim1$ ps, corresponds to the beat between the pair of bright modes separated by $\sim$ 35 cm$^{-1}$ (Fig. 1d).  In the presented case in Fig. 2c, the signal suddenly drops. The particle becomes SERS inactive, while its TEM shows an intact dumbbell (Fig. 2b, inset). The measurement sequence suggests that the loss of signal is due to the separation between molecule and hot spot.  This common occurrence limits the tolerable exposure time, therefore sampling statistics, of measurements.

Significant variation is seen in tr-CARS signals recorded on individual dumbbells. Examples are shown in Fig. 3a. When all three pulses coincide, at $\tau$ = 0, dramatic interferometric fluctuations are observed, which are blanked out in the figure. At negative delay, when the probe precedes the preparation pulses, the time independent two-beam electronic CARS signal of the metal antenna is seen. The fluctuations at negative time establish the noise floor over which the molecular response appears upon passing $\tau$ = 0 (Fig. 3a, orange and gray curves). No identifiable molecular response is seen from the SERS inactive structure (Fig. 3a, green curve). Modulation with identifiable periodicity is only seen at positive time, and only on SERS active particles. The CARS signal shows identical features when simultaneously detected in the forward (gray curve) and backward (orange curve) scattering directions, indicating that the oscillation of the signal is an intrinsic response of the dumbbell-molecule system, independent of detector noise. These signals were acquired with incident peak intensities that reach $\sim10^{10}$ W/cm$^2$. Under these irradiation conditions, we find surprisingly little change to the dumbbell. Detailed TEM imaging reveals that pre-irradiated dumbbells show a clear junction gap of $< 2$ nm between nano-spheres (Fig. 3b), while post irradiation images show the formation of a neck between spheres (see Fig. 3c,d). Although it is unclear at what stage of the illumination this morphological change occurs, the dumbbells remain largely intact and the CARS signal remains remarkably stable during the 1 hr duration of the measurement. 

Examples of time traces that are characteristic to tr-CARS in the single BPE molecule limit are shown in Fig. 4, along with the signal of the ensemble of solid BPE recorded under identical conditions. The contrast is striking. The ensemble signal decays in 1 ps (Fig. 4a), while the depth of modulation of the single dumbbell signals persists for the 10 ps duration of the measurement (Fig. 4b,c). The ensemble signal, given by the third-order polarization\cite{Mukamel1995}:
\begin{equation}
S(\tau)=\int dt\left|\left<P^{(3)}(\tau)\right>\right|^2
\label{eq:pol}
\end{equation}
reduces to the damped quantum beats of the prepared vibrational coherence:
\begin{equation}
S(\tau)\propto\sum_{\nu,\nu'}a_{\nu}a_{\nu'}\cos(\omega_\nu-\omega_{\nu'})e^{-(\gamma_\nu+\gamma_{\nu'})\tau}
\label{eq:ens}
\end{equation}
with phenomenological dephasing rates $\gamma_{\nu,\nu'}$ that arise from the ensemble averaging indicated by the angle brackets in (\ref{eq:pol}).  The same information is contained in the Raman spectrum of the bulk (Fig. 4a inset), which in the absence of electronic resonances, is given by the same third-order response. \cite{Mukamel1995} Consequently, the tr-CARS signal can be retrieved by the windowed Fourier transform of the Raman spectrum, as shown in Figure 4a. The broad Lorentzian profile of the Raman lines leads to rapid exponential dephasing of the signal in time. This is clearly not the case for signals recorded on the dumbbell. Given the much broader SERS spectrum (Fig. 4a, inset), the CARS signal would have been expected to decay even faster than the ensemble, on 0.3 ps time scale. Instead, the depth of modulation of the tr-CARS signal from the dumbbell (Fig. 4b) persists for the duration of the measured time delay $\tau$ = 10 ps.  A vibrational superposition of four states, given by the stick spectrum in Fig. 4a, emulates the signal with good fidelity. The main beat is that of the two breathing modes separated by 35 cm$^{-1}$ $(\omega$ = 1647 cm$^{-1}$, 1612 cm$^{-1}$). The higher frequency beat is reproduced by the inclusion of lines at $\omega$ = 1580 cm$^{-1}$, which are seen in the transient SERS spectrum from the same dumbbell (Fig. 4a, inset); and the splitting between the two modes of 5 cm$^{-1}$ reproduces the slow modulation of the signal envelope.  The main features of the signal are satisfactorily reproduced. Although alternate interpretations may be given, what is clear is that there is no evidence of dephasing. If more than one molecule contributes to the signal, then their collective spectral distribution must remain within $\sim$ 0.3 cm$^{-1}$ for the 1 hr duration of the measurement. This seems unlikely. The signal is best described as tr-CARS in the single molecule limit. 

To be more definitive about the single molecule origin of a signal, it is necessary to track the wavepacket motion until the onset of quantifiable noise.  An example is provided by the tr-SECARS signal of Fig. 4c. Note that each experimental point of the time trace is the average of 50 measurements, taken with $\sim2 \mu$s dwell time per pixel at 1 s intervals; and we estimate a total of 5 - 10 signal photons detected per point. The total accumulation time is $\sim$ 1 hr for the full trace. Spectral fluctuations during the measurement intervals generate phase noise. Thus, the discretely sampled normalized signal at a given delay time $\tau$ can be generally described as:
\begin{equation}
S(\tau)_{n,m}=\frac{1}{M}\sum^{M}_{m=1}\left|\frac{1}{N}\sum^N_{n=1}\frac{1}{V}\sum^V_{\nu=1}a_\nu e^{-i(\left<\omega_\nu\right>+\delta_{\nu,m,n})\tau}\right|^2
\label{eq:single}
\end{equation}
in which $M$ is the number of realized measurements (photons), $N$ is the number of molecules, $\left<\omega_\nu\right>$ is the mean wavenumber of vibrational mode $\nu$ and $\delta_\nu$ is its stochastic fluctuation around the mean. As suggested by the Gaussian profiles of SERS lines (Fig. 1), $\delta_\nu$ derives from a normal distribution $N(0,\sigma)$ with zero mean and variance $\sigma$  given by the spectral width of the vibrational lines. The phase fluctuation grows with delay time $\tau$, such that at $\sigma\tau >> 2\pi$ the random phase $\delta\tau$ is uniformly distributed over the full $\left[-\pi,\pi\right]$ interval. In this limit, the probability distribution of observable signal amplitudes reduces to a normal distribution, with a characteristic mean: 
\begin{equation}
\left<S(\tau>2\pi/\sigma)\right> =\frac{1}{N V}
\end{equation}
It is the summation over the coherence terms inside the square in (\ref{eq:single}), over $N$ molecules and $V$ modes, that leads to dephasing: $\left<S(\tau) \right> \rightarrow 0$ for large $N$ and $V$. For a single molecule and a superposition of two states, the expected value is:
\begin{equation}
\left<S(\tau>\pi/\sigma)\right>=\frac{1}{2}\;  \mbox{for}\; N=1,V=2
\end{equation}
Note, in a given measurement, the stochastic phase leads to a random value of the normalized sinusoidal signal $S(\tau)\in\left[1,0\right]$, with mean $\left<S(
\tau)\right>=1/2$ and variance $1/\sqrt{M}$ given by averaging over the $M$ realizations. This assumes no correlation in spectral fluctuations $\delta_\nu$ between different modes. In the other extreme, assuming fully correlated fluctuations, the signal retains full coherence. This can be clearly seen by considering the superposition of two states on a single molecule:
\begin{equation}
S(\tau)=\frac{1}{2}+\frac{1}{2M}\sum^M_{m=1}\cos(\left<\omega_\nu-\omega_{\nu'}\right>\tau+(\delta_\nu-\delta_{\nu'})_m\tau)
\label{eq:twostates}
\end{equation}
If in each realization, $m$, the fluctuation is correlated, $\delta_\nu-\delta_{\nu'}=0$, the signal does not dephase. 

More generally, in an open system, spectral fluctuations are stochastic. The expected signal for a single quantum beat $(V=2)$, on a single molecule $(N=1)$, and $M = 5$ realizations sampled from a normal distribution $(\delta_\nu-\delta_{\nu'})\in\mathcal{N}(0, \sigma = 1.5$ cm$^{-1})$ is illustrated in Fig. 4c (green curve). The assumption of two molecules produces a different trace at long delays, since in this case $\left<S(\tau>\pi/\sigma)\right>\rightarrow1/4$ (Fig. 4d).  Despite the stochastic nature of the trajectories, it is possible to assign the prepared superposition to the vibrational coherence of the two states that dominate the ensemble response, albeit with a spectral distribution that is significantly narrower than observed SERS linewidths. We learn that there is significant correlation between the two C=C modes. This highlights an important difference between CARS and Raman that becomes pronounced in the single molecule limit. Through the squared third-order polarization in Eq. (\ref{eq:pol}), CARS measures the cross-correlation of fluctuations between modes (\ref{eq:twostates}), while linewidths in Raman measure the dephasing of individual modes.

Stochastic trajectories can be rigorously  compared through their statistics. Unique assignments are possible through the probability distribution functions (PDF), namely, the histograms of the time traces shown in the right panel of Figure 4. In the evolution interval $0<\tau<2\pi/\sigma$, the observable distributions and the expected values in signal amplitude are oscillatory functions that obey phasor statistics.\cite{Goodman2007} The characteristic PDFs can be generated numerically, as shown in Fig 5. By inspection, the histogram of the signal (of the time trace in Fig. 4c) can be uniquely assigned to a single vibrational beat of a single molecule. Moreover, based on the indicated first two moments of the PDFs, the spectral co-variance, $\sigma= 1.5$ cm$^{-1}$, can be directly bracketed. The comparison between experimental PDF and the numerical characteristic distributions can be quantified through the nonparametric Kolmogorov-Smirnov (KS) test.\cite{Young1977} Based on the KS-distance between cumulative distributions, it is possible to establish that the PDF ($N=1,V=2,\sigma=1.5$ cm$^{-1}$) is the best match to the experiment, and that the experiment obeys the single molecule, single beat statistics with 99\% likelihood. More generally, the figure illustrates that the information content in time dependent measurements of coherences in the single molecule limit is contained in the correlated probability densities of phase and amplitude variations. While Eq. (\ref{eq:single}) yields to rigorous statistical analysis, the principle of key relevance to the present is that a single molecule cannot dephase but develops characteristic phase noise. In contrast, multiple molecules in a multi-mode coherence rapidly dephase (see Fig. 5). The early time PDF of the signal yields a statistically meaningful distinction between one versus two versus many molecules when a small number of states is prepared and interrogated. Experimentally, we observe all possible variations: some traces reflect single molecule behavior, others indicate the presence of a few to many molecules. Examples are shown in the supplementary material, along with their KS analysis. 

In summary, through the persistent vibrational coherence on a single nano-antenna (Fig. 4b,c) we presented the first real-time record of vibrational wavepacket motion at the single molecule level, under ambient conditions. To reach single molecule sensitivity, we employed the surface enhanced Raman effect in the time domain, at the plasmonic junction of gold nanospheres. We demonstrated the viability of antennaed molecules under ultrafast irradiation, and introduced time-resolved, surface enhanced CARS as a tool to interrogate molecular motions in the single molecule limit.  We should note that as a four-wave mixing (FWM) process, CARS belongs to one of the more flexible multi-dimensional NLO schemes.\cite{Mukamel1995} It has the attributes of Raman scattering in its generality, and under SERS conditions, dramatic E$^8$-enhancements are to be expected. As a coherent scattering process, tr-CARS allows the selective preparation and interrogation of superposition states, with time resolution limited by the laser pulses. Complete quantum state reconstruction is possible through PDFs in electronically resonant FWM,\cite{Golschleger2013} and it has previously been recognized that quantum logic can be mapped on the preparation, evolution and measurement steps of FWM.\cite{Zadoyan2001, Bihary2002, Glenn2006} The manipulation of truly quantum information becomes accessible in the single molecule limit. The present experimental demonstration of coherent wave-packets sustained on single molecules, paves the way toward such applications. 

\begin{acknowledgments}
The authors wish to thank R. P. Van Duyne for providing us the samples, P. Z. El-Khoury for DFT calculation of BPE, and N. Apkarian for pointing out the KS analysis. SEM and TEM work was performed at the Laboratory for Electron and X-ray Instrumentation (LEXI) at UC Irvine. This work is made possible by the National Science Foundation Center for Chemical Innovation on Chemistry at the Space-Time limit under grant CHE-0802913. E.H. is supported by the Academy of Finland Decision No. 265502.
\end{acknowledgments}

\noindent\textbf{Author Contributions.}\\
SY and DAF conducted the TR-CARS measurements; MB, EH and SD carried out the Raman and EM measurements; SY and VAA performed the analysis; EOP and VAA conceived of the work and wrote the manuscript.\\
\noindent\textbf{Competing financial interests}\\
The authors declare no competing financial interests.
\pagebreak
%\bibliography{your-bib-file}

\begin{figure}[h]
\centering
\includegraphics[width=12cm]{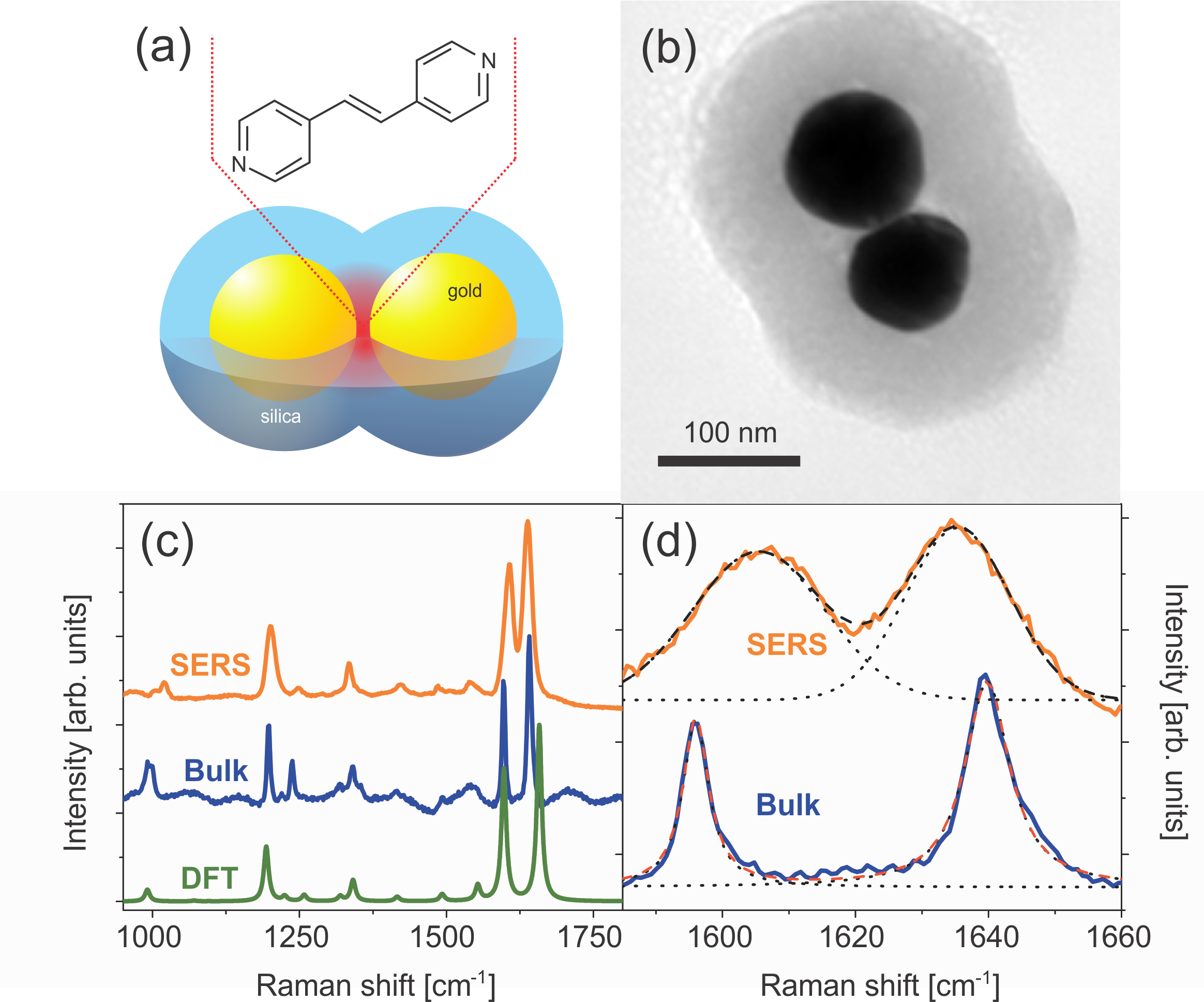}
\caption{(a) Sketch of the dumbbell-shaped gold nano-structures with adhered BPE molecules and encapsulated in a silica shell. Strong SERS response is expected from BPE molecules in the hotspot between the gold nano-spheres. (b) TEM of a single dumbbell structure. (c) Comparison of SERS spectrum from a single dumbbell, Raman spectrum from bulk, and DFT calculation of the Raman spectrum of BPE. (d) Comparison of SERS spectrum and bulk Raman spectrum in the vicinity of 1600 cm$^{-1}$.}
\end{figure}
\pagebreak
\begin{figure}[h]
\centering
\includegraphics[width=12cm]{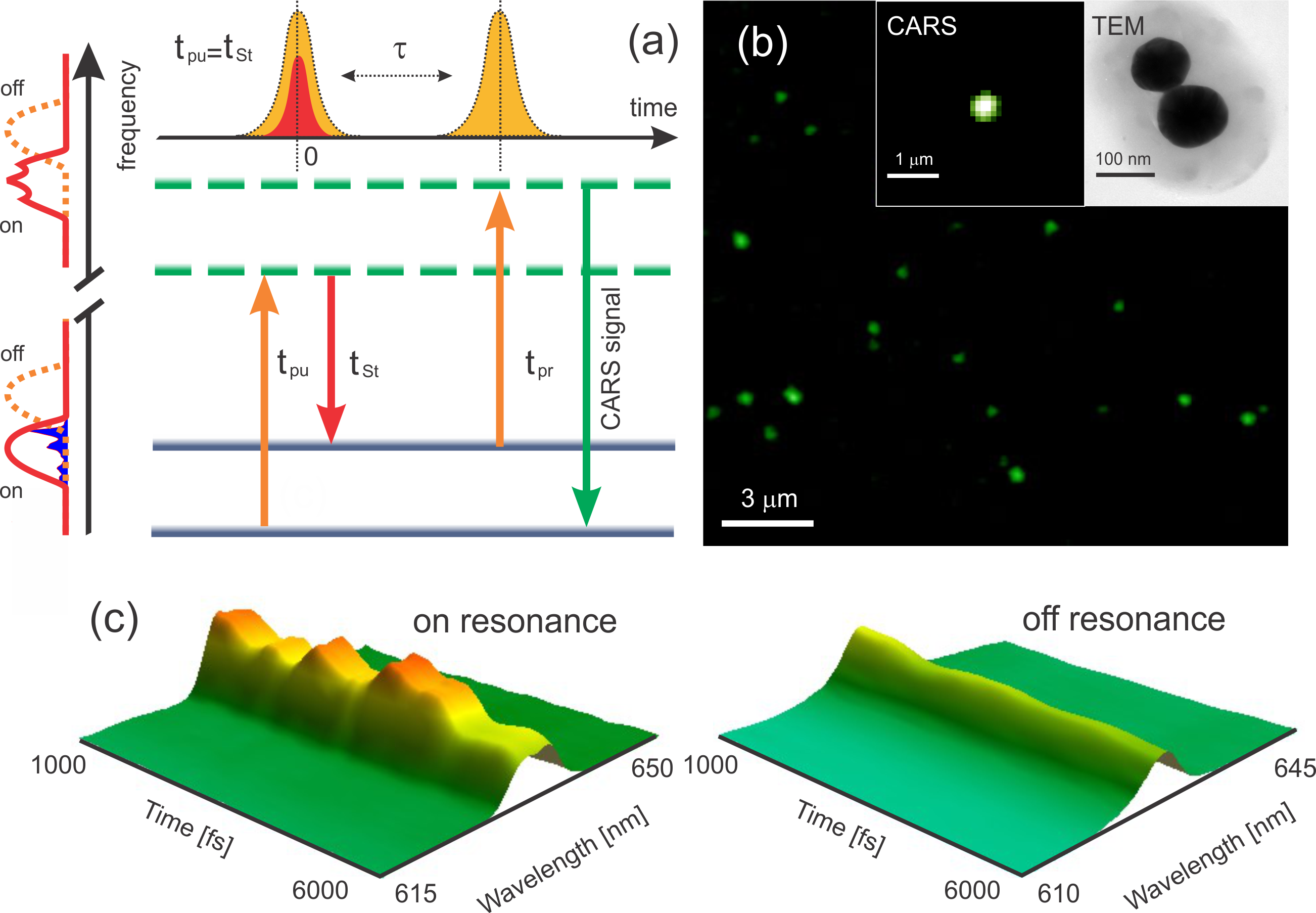}
\caption{(a) Jablonski diagram of the tr-CARS process, with the time delay $\tau$ between the pump/Stokes pair and the probe pulse indicated. (b) CARS image of isolated dumbbell structures. The inset shows the CARS image of a single dumbbell and the corresponding TEM image. (c) Spectrally resolved tr-CARS of a single dumbbell acquired on resonance at $\omega_{pu}-\omega_{St}=1640$ cm$^{-1}$ (left) and acquired off-resonance at $\omega_{pu}-\omega_{St}=1800$ cm$^{-1}$ (right). Note that the clear quantum beats disappear when tuned off-resonance.}
\end{figure}
\pagebreak
\begin{figure}[h]
\centering
\includegraphics[width=12cm]{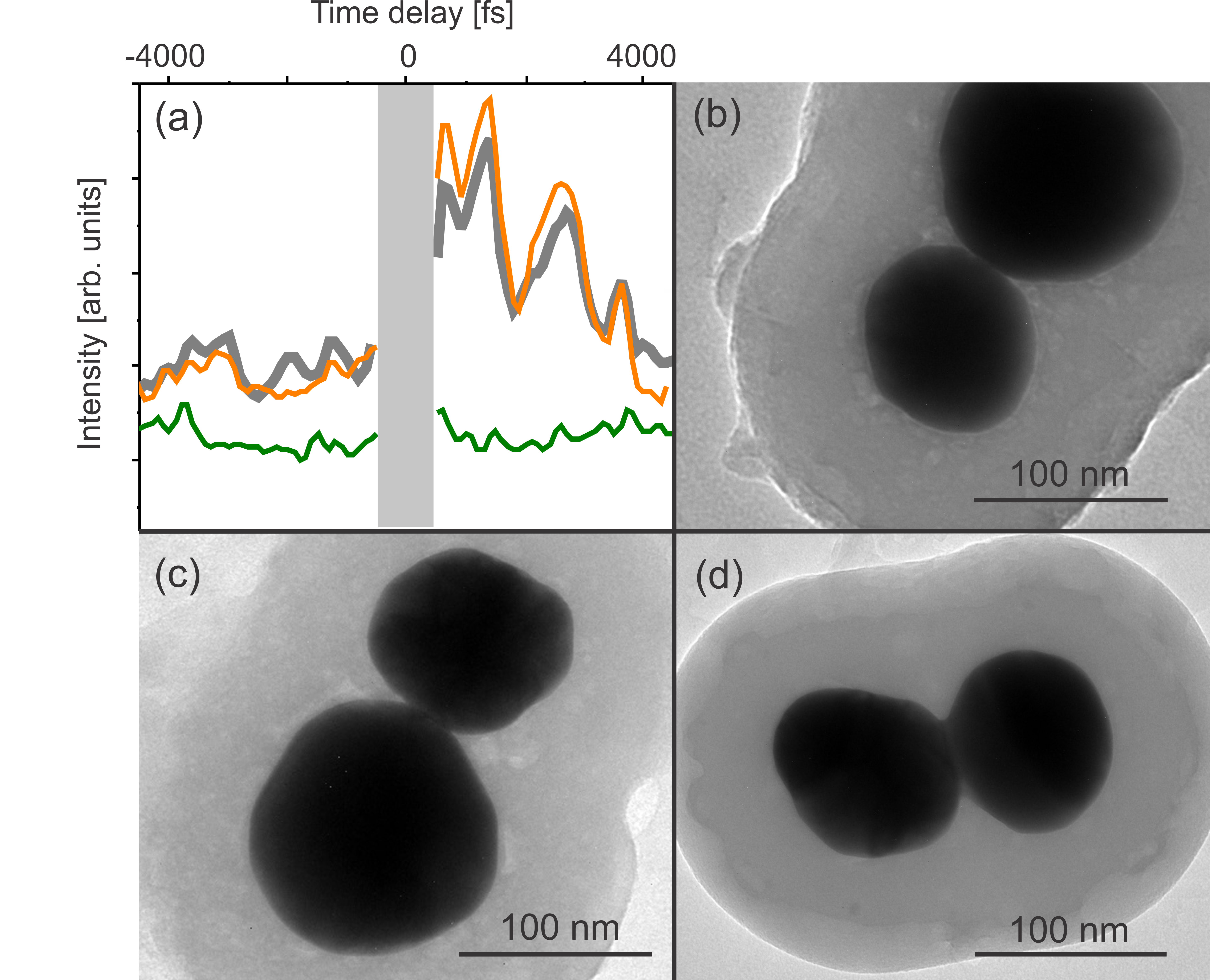}
\caption{(a) tr-CARS of individual dumbbell structures, including a SERS-active structure (orange and gray curve), and a SERS-inactive structure (green curve). The orange curve is the back-scattered CARS and the gray curve is the simultaneously measured forward scattered CARS. (b) TEM of a dumbbell before optical illumination shows a clear gap between the two spheres. (c,d) Examples of TEM images of dumbbells  after SERS optical experiments show formation of a neck between the spheres.}
\end{figure}
\pagebreak
\begin{figure}[h]
\centering
\includegraphics[width=12cm]{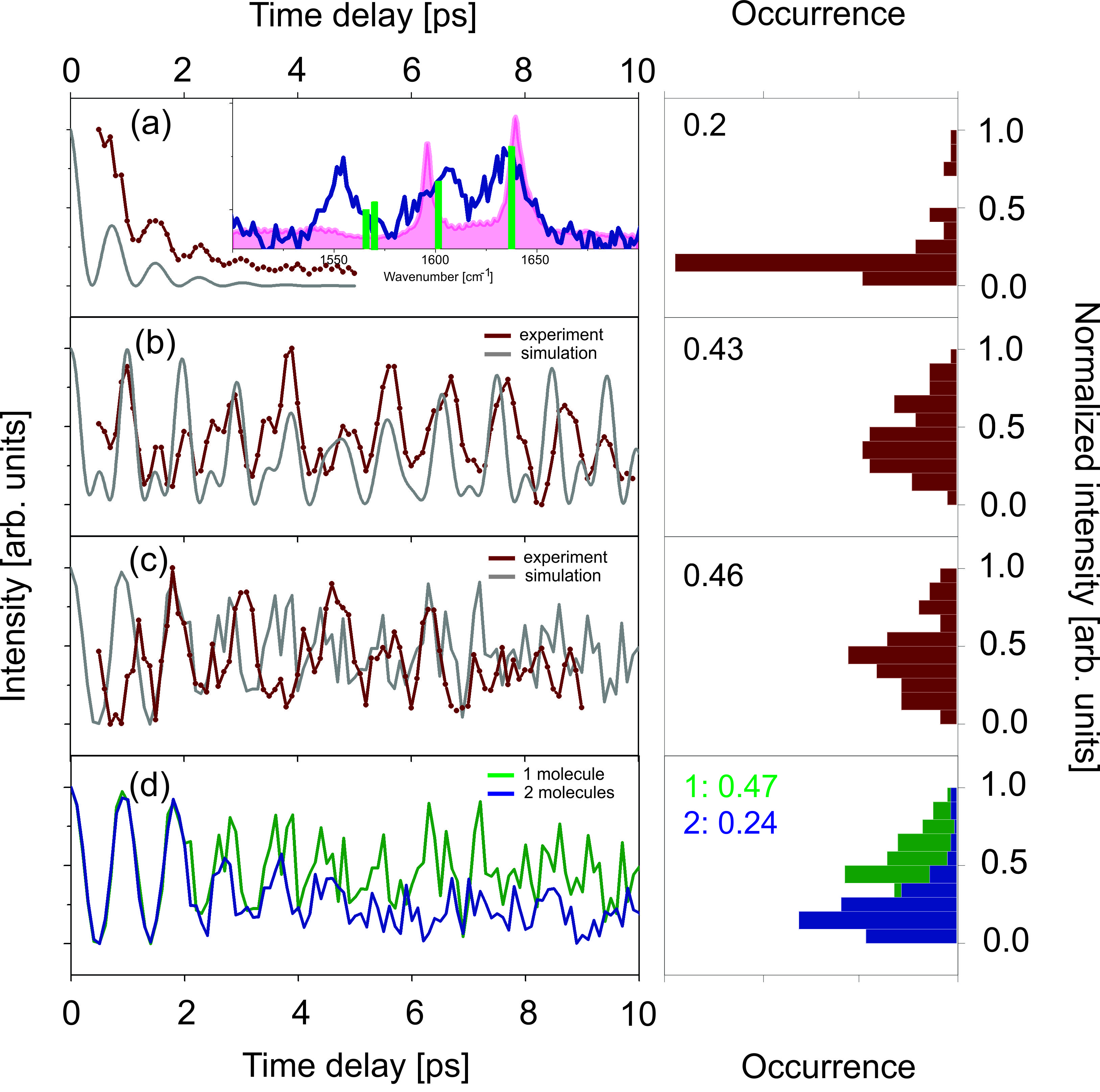}
\caption{(a) tr-CARS of bulk BPE (brown curve) along with the windowed Fourier transform of the bulk Raman spectrum (gray curve). The inset shows the bulk Raman spectrum (pink) and a transient SERS spectrum recorded on the same dumbbell on which tr-CARS trace (b) was obtained. (b) tr-CARS trace of a single structure (brown curve) shows distinct quantum beats. The gray curve is a simulation of the tr-CARS response based on the stick spectrum in the inset of (a). (c) tr-CARS of another isolated nano-structure (brown curve) showing  periodic oscillations at early time followed by phase noise at longer time. The gray curve is a simulation based on equation (\ref{eq:twostates}) for a single molecule with $M=5$ and $\sigma=1.5$ cm$^{-1}$. (d) Comparison between one-molecule (green) and two-molecule (blue) simulated signals. The right panels show the PDF and their first moments derived from the corresponding temporal traces in the panels to the left.}
\end{figure}
\begin{figure}[h]
\centering
\includegraphics[width=14cm]{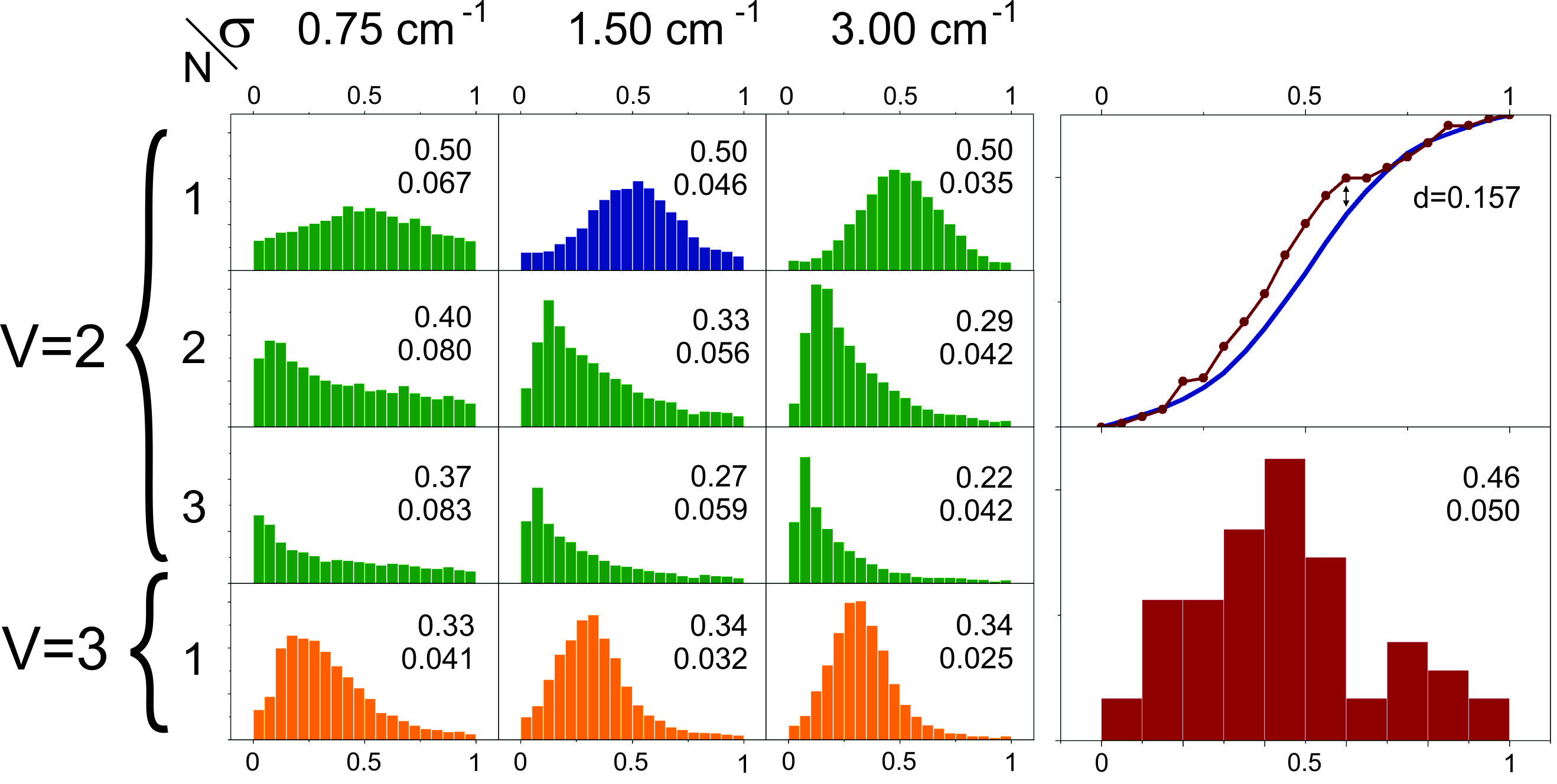}
\caption{Probability distribution functions (PDF) calculated for different realizations of the number of molecules $(N)$, the number of modes $(V)$ and the variance of the spectral fluctuations ($\sigma$). The first two moments of the distributions are indicated in each panel. The panel to the right gives the PDF for the tr-CARS trace shown in Figure 4c. The experimental PDF (brown bars), and its two moments, uniquely match the simulated PDF for $N=1$, $V=2$ and $\sigma=1.50$ cm$^{-1}$ (blue bars). The experimental cumulative distribution function (CDF; brown dots, top right) and the theoretical CDF for the simulated PDF (blue line) are also indicated. The Kolmogorov-Smirnov (KS) test reveals a a maximum distance of $d=0.157$ between the experimental and theoretical CDFs, translating into a $99\%$ likelihood that the data represents the evolution of a statistical two-state superposition on a single molecule.}
\end{figure}
\end{document}